\documentclass[12pt]{article}
\usepackage{amsfonts,amsmath,amssymb}
\usepackage{scalerel}[2016/12/29]
\usepackage{graphicx}
\usepackage{tikz}
\usepackage{braket}
\usepackage[nosort]{cite}

\usepackage{hyperref}

\usetikzlibrary{arrows}
\usetikzlibrary{decorations.markings}

\newcommand{\nn}{{\nonumber}}

\def\be{\begin{equation}}
\def\ee{\end{equation}}
\def\bea{\begin{eqnarray}}
\def\eea{\end{eqnarray}}

\def\half{{\textstyle {1\over 2}}}

\def\bb{\hskip -0.5mm}
\def\bbb{\hskip -0.75mm}

\def\hb{\hskip -0.25mm}

\usepackage{amsmath}
\usepackage{epsfig,graphicx}
\usepackage{graphicx}
\usepackage{tikz}

\def\be{\begin{equation}}
\def\ee{\end{equation}}
\def\bea{\begin{eqnarray}}
\def\eea{\end{eqnarray}}

\begin{document}

\thispagestyle{plain}

\title{\bf\large Quantum mechanical rules for observed observers and the consistency of quantum theory}

\author{Alexios P. Polychronakos$^{1,2}$}

\date{March 22, 2024}

\maketitle

%

{\em
{\centerline{$^1$Physics Department, the City College of New York, NY 10031, USA}}
\vskip .2 cm
{\centerline{$^2$The Graduate Center, CUNY, New York, NY 10016, USA}
\vskip .3 cm
\centerline{\it apolychronakos@ccny.cuny.edu}
}}

\maketitle

\begin{abstract}
I argue that the rules of unitary quantum mechanics imply that observers who will themselves be subject to
measurements in a linear combination of macroscopic states (``cat" measurements) cannot
make reliable predictions on the results of experiments performed after such measurements. This lifts the inconsistency
in the interpretation of quantum mechanics recently identified by Frauchiger and Renner.
The Born rules for calculating the probability of outcomes and for communicating with other observers
do not generally apply for cat-measured observers, nor can they generally be amended to incorporate upcoming
cat measurements. Quantum mechanical rules completed with these conditions become fully consistent.
\end{abstract}

The interpretation of quantum mechanics in the context of measurements, and concepts such as state ``collapse," have
troubled physicists since the inception of quantum theory. Pushed to their logical extreme, such issues
become entangled (in the colloquial sense) with questions of consciousness, reality, etc. Bell's theorem eliminated
possible ways out of this tangle via any (halfway reasonable) classical underlying theory, and the success of
quantum mechanics forces us to adopt it as a fundamental theory and face the logical consequences.

The crux of the matter is the privileged role of the observer in quantum theory. Every interpretation of quantum mechanics
is formulated in terms of what an observer expects to see or measure (any other formulation would be unscientific).
This dichotomy between observer and (quantum) system raises some obvious questions: are observers subject to the laws of
quantum mechanics? (Obvious answer: yes; otherwise the theory would be openly incomplete.) And, who is an observer?
(Plausible and possibly incomplete answer:
any sufficiently complicated macroscopic system.)

Accepting that quantum theory applies to observers, the isue of what happens to them if they are themselves observed
by another (super?)observer becomes important. The question goes back to Schr\"odinger and his unfortunate cat, and was sharpened by
a treatise of Wigner \cite{Wig} which promoted the cat to another conscious observer, in what is nowadays called the ``Wigner's friend" setup. Still, Wigner's thesis neither resolved the question of measurements on observers nor identified any
inconsistencies in quantum mechanics and its traditional interpretation.

Recently, Frauchiger and Renner (FR) proposed a thought experiment \cite{FR} built on a
variant of the ``Wigner's friend'' setup paralleling a construction by Hardy \cite{Hardy} that, upon application of the conventional rules of quantum mechanics, leads to a
contradiction, thus casting doubt on the logical consistency of quantum theory when its application is extended to
observers themselves. The FR argument crucially relies on a situation in which an observer is ``measured''
with respect to a linear superposition of macroscopically distinct states. Then, application of conventional Born
rules and consistency between observers lead to the contradiction. As expected, the arguments and conclusion of FR have
become the object of much commentary and debate, and various explanations, remediations, and (often sharply worded)
criticism have been offered (\hskip-0.02cm\cite{Aron,Arau,LH, Kastner,many,manymore,talk} is a small, incomplete sample).

In this note I point out that quantum mechanics requires (in fact, implies) that when such
macroscopic measurements happen on observers, then these observers cannot use the standard Born rules of
quantum mechanics to predict the results of measurements that will be completed after they suffer such a measurement.
In effect, this extends the quantum mechanical mantra ``measurements disturb the system'' to also apply to observers. Communication of information between such observers is similarly affected.

The argument presented here is applicable to any processes where such measurements are performed and to any deduction
based on the predictions of observers for an event after they have suffered such a measurement, implying 
that conclusions based upon such deductions are not warranted. I will, however, give a more detailed explanation of
how the argument applies to the FR thought experiment, which served as the original motivation for this work, and demonstrate
that the completion of the quantum mechanical rules proposed here eliminates the inconsistency.
I stress that this is not a
``refutation" of FR's argument, which solidly relies on a set of assumptions and remains logically valid. It rather complements FR's work
by making a (required by quantum mechanics, I argue) modification in their assumptions, which lifts the inconsistency.

{{\bf Basic approach}}: I start by declaring that the approach taken in this work relies on
strictly unitary evolution of states and on the standard definition of measurement as entanglement
between observer and observed system \cite{Hepp}.
Unitary evolution is the only one compatible with relativistic quantum mechanics and quantum field theory,
and, upon proper interpretation, can account for all observed phenomena.

In this approach, there is no
fundamental concept of state collapse, and the ensuing certainty of observers about their observation
outcomes is encoded in the entanglement between their state and the state of the measured system.
After the measurement, the full state becomes a superposition of orthogonal components, each
consisting of the state of the observer having observed a particular outcome entangled with the
eigenstate of the measured system in that outcome. The full system still represents an observer
being certain about the outcome of the measurement, as each orthogonal component shares this
property (we may say that the full state is an eigenstate of the ``certainty operator").

This approach is essentially equivalent to a ``many worlds" interpretation, as each orthogonal component
of the state can be considered as a different ``branch" of the universe. This interpretation is consistent
but not strictly needed: under normal circumstances the orthogonal components are `superselected';
that is, no transition between them can occur, and thus no physical process can reveal the presence of
other branches to an observer in one of them. Other branches are, therefore, epistemologically irrelevant.
However, measurements of observers in linear combinations of macroscopic states
(``cat" measurements) explicitly induce transitions between the branches, lifting the superselection
property and rendering
the many worlds interpretation less useful (one may say that branches of the world recombine and mix).

At any rate, the unitary evolution approach is incompatible with other alternative interpretations,
such as QBism, and I will have nothing to say about such interpretations.

{{\bf The basic argument}}: The argument will be formulated in terms of pure states, but can
easily be extended to ensembles of states (density matrices). 

Consider the Wigner's friend situation in its simplest: a system consisting of a spin-half $S$ and two observers $A$ and $\rm B$ (whom, since they engage in types of measurements in which Alice and Bob never did,
I prefer to think of as Alex and Barbara).
$A$ can perform measurements on $S$, but $\rm B$ can also perform measurements on $A$, and with
respect to states that are superpositions of distinct
macroscopic (cognitive) states of $A$. I will call such states ``cat'' states (a standard term), and such
measurements ``cat" measurements.

Initially, the system is in a pure unentangled state $\ket{S} \ket{A} \ket{\rm B}$ (tensor products $\otimes$ are understood).
We work in the Schr\"odinger representation and assume, for simplicity, that states evolve only when they interact. The process we consider
is represented by the state evolution shown below ($\ket{\uparrow}$ and $\ket{\downarrow}$ are the
standard $z$-axis spin eigenstates):
{\small{\bea
& \text{Initial~state} ~~~~~~~~~&~~~~\textstyle{1\over \sqrt 2} \bigl(\ket{\uparrow} + \ket{\downarrow} \bigr) \ket{A} \ket{\rm B} \\
~ &~ \cr
&\hskip -0.7cm A~\text{measures~spin~in~$z$~axis}~\Rightarrow ~~~&~~\textstyle{1\over \sqrt 2} \Bigl(\ket{\uparrow} \ket{U} + \ket{\downarrow} \ket{D} \Bigr) \ket{\rm B} \\
& ~~         &\bb\bb=\textstyle{1\over \sqrt 8} \Bigl\{ \ket{\uparrow} \bb\bigl[(\ket{U}\bb+\bb \ket{D})\bb +\bb (\ket{U} \bb-\bb \ket{D})\bigr]  
+ \ket{\downarrow} \bb\bigl[ (\ket{U}\bb+\bb \ket{D})\bb -\bb (\ket{U}\bb-\bb \ket{D}) \bigr] \Bigr\} \ket{\rm B}\nn \\
~ &~ \cr
&\hskip -1.1cm {\rm B}~\text{measures~{\it A}~in~cat~state}~\bb\Rightarrow &\hskip -0.2cm \textstyle{1\over \sqrt 8} 
\Bigl\{ \ket{\uparrow} \bigl[(\ket U\bb\bb+\bb\bb \ket D)\bb \ket{\rm Y}\bb+\bb(\ket U \bb\bb-\bb\bb \ket D)\bb \ket{\rm N}\bigr]  
+ \ket\downarrow \bigl[ (\ket U\bb\bb+\bb\bb \ket D)\bb\ket {\rm Y}\bb -\bb (\ket U\bb\bb-\bb\bb \ket D) \bb\ket{\rm N} \bigr] \Bigr\}\nn \\
&~~~ & =\textstyle{1\over \sqrt 8}  \ket U\Bigl( \ket\uparrow\bb \ket{\rm Y}\bb+\bb \ket\uparrow \ket{\rm N} \bb+\bb\ket\downarrow \bb\ket{\rm Y}\bb -\bb \ket\downarrow \ket{\rm N} \Bigr)\bb  \label{Cat} \\
&~~~ &+\,\textstyle{1\over \sqrt 8} \ket D \Bigl(\ket\uparrow\bb \ket{\rm Y}\bb - \bb\ket\uparrow \ket{\rm N}\bb +\bb\ket\downarrow \bb\ket{\rm Y}\bb +\bb \ket\downarrow \ket{\rm N} \Bigr)~~~
\nonumber
\eea}}
Initially, the spin is set to the state $\ket\rightarrow = \bigl(\ket\uparrow + \ket\downarrow\bigr)/\sqrt{2}$. At some time, 
observer $A$ measures the spin in the $z$-axis.
After the measurement, the states of $A$ and $S$ become entangled, with state $\ket U$, representing $A$ having
observed an up-spin, entangled with $\ket\uparrow$, and state $\ket D$, representing $A$ having observed a down-spin,
entangled with
$\ket\downarrow$. After that, the spin is left alone and is not touched by anyone. If $A$ were to measure the spin again,
$U$ would definitely find it to be up and $D$ would definitely find it to be down. 

At a later time, observer $\rm B$ performs a cat measurement on $A$. Specifically, $\rm B$ checks if
observer $A$ is in the cat state $\ket U\bb+\bb\ket D$, entangling a state $\ket{\rm Y}$ representing $\rm B$ having given the
answer Yes with the state $\ket U\bb+\bb\ket D$, and a state $\ket{\rm N}$ of $\rm B$ having given the answer 
No with the orthogonal state $\ket U\bb-\bb\ket D$. The final state is as in (\ref{Cat}).

Assume, now, that observer $A$ measures the spin again. The results will be either up or down,
irrespective of the value observed previously. If $A$ had originally found the spin to be up,
he now has a 50\% chance of finding it down. And yet nobody had touched the spin!
What has happened is that the observer himself was touched and measured, in a dramatic way that
altered the entanglement of his cognitive state with the observed state of the spin.

The lesson we draw from this is:
\vskip -0.1cm
\noindent
{\it Observation 1:} Observers cannot in general apply the standard Born probability rules if
they themselves will be subject to cat measurements.

Observers, of course, do not know the full state of the universe, and often not even the full state of their
environment. In general, they know the state of part of their system and update this knowledge as they
gather information from measurements they perform or interactions with other observers. This is so,
in particular, in the original Wigner's friend setup. For situations not containing
cat measurements, deductions based on such partial states are consistent with deductions based on the
full state of the system, differing only in the degree of their predictability. Crucially, this is not the case
in situations involving cat measurements, and this
is the essence of Observation 1 above. To make this explicit, we analyze the situation of eq.~\ref{Cat}
in the context of the states perceived by each observer.

Assume, for concreteness,
that $A$ and $\rm B$ know nothing initially about the state of the system. $A$ is only aware of the presence
of the spin, while $\rm B$ is only aware of the presence of $A$ (their respective measured systems) and,
of course, both know their own state. The initial states assumed by each observer are
\bea
&&~\text{for~}A \hskip 4.9cm \text{for~}{\rm B} \cr
&&\ket A \ket{S'}  \hskip 4.4cm \ket {\rm B} \ket {A'} \label{reda}
\eea
where $\ket{S'}$ and $\ket{A'}$ are generic unknown states for the spin and $A$.
After $A$ performs the measurement of the spin, the updated states are
\bea
&&~\text{for~}A \hskip 6.8cm \text{for~}{\rm B} \cr
&&\ket U \ket \uparrow ,~\text{if~up~was~observed}   \hskip 2.8cm \ket {\rm B} \ket {A'}\\
&&\ket D \ket \downarrow ,~\text{if~down~was~observed}\nonumber
\eea
$A$ can conclude at this point that, if the spin remains undisturbed and he performs
his spin measurement again, the probabilities of the outcomes based on his present state are $100\%$
to find the spin up if it was up before, and $100\%$ down if it was down before.
After $\rm B$ measures $A$, the updated states are
\bea
&&~\text{for~}A \hskip 5.5cm \text{for~}{\rm B} \cr
&&\ket U \ket \uparrow ,~\text{if~up~was~observed}   \hskip 1.5cm \ket {\rm Y} (\ket U + \ket D)/\textstyle{\sqrt 2} ~,~\text{if~Yes~was~observed} \cr
&&\ket D \ket \downarrow ,~\text{if~down~was~observed} \hskip 1cm \ket {\rm N} (\ket U - \ket D)/\textstyle{\sqrt 2} ~,~\text{if~No~was~observed}
\label{penu}\eea
Finally, $A$ performs his second measurement of the spin and the updated states are
\bea
&&~\text{for~}A \hskip 7cm \text{for~}{\rm B} \cr
&&\ket {UU} \ket \uparrow ,~\text{if~up~\&~then~up~observed} \hskip 1.5cm \ket {\rm Y} (\ket U + \ket D)/\textstyle{\sqrt 2} ~,~\text{if~Yes~observed} \cr
&&\ket {UD} \ket \downarrow ,~\text{if~up~\&~then~down~observed} 
 \hskip 1cm \ket {\rm N} (\ket U - \ket D)/\textstyle{\sqrt 2} ~,~\text{if~No~observed}\cr
&&\ket {DU} \ket \uparrow ,~\text{if~down~\&~then~up~observed} \cr
&&\ket {DD} \ket \downarrow ,~\text{if~down~\&~then~down~observed} \label{redw}
\eea
The two middle outcomes in $A$'s state should not have occurred according to his predictions based
on his states at (\ref{penu}). Yet they do occur, according to the full unitary evolution of the system,
and violate $A$'s predictions. $A$ might be tempted to conclude that the spin was disturbed,
but this is not a justified conclusion: $A$ could have made sure that the
spin was isolated and protected from external influences. The only conclusion that $A$ can draw, then,
is that his application of Born rules provided unreliable results.

Are $A$'s unreliable predictions due to his brain having somehow been ``scrambled"
by the cat measurement? Is $A$ even aware that he has been cat measured? In fact, I would argue that neither
is true: in a ``clean" cat measurement (involving the minimal measuring operator) the thought process of $A$
is not disturbed. This will be demonstrated later, when the execution and feasibility of cat measurements
are examined. At any rate, the effects of a cat measurement on the observer's conscious state and the full
details of quantum (cat) vs. classical meddlings with the observer's mind are open to
interpretation and might be an issue worth exploring in the future.

Could perhaps observer $A$ modify his application of quantum mechanical rules to account for
measurements that he knows will happen to him? Sadly, in general no. To do so, $A$ should know
the exact state of the full system before he performs any measurements, as well as the precise
measurement that will be performed on him afterwards. With anything short of this full information,
$A$ can make no reliable predictions, even probabilistic ones.

As a demonstration, consider that $A$ has no knowledge of the spin state before he measures it, but knows of
the presence of $\rm B$ and what exactly she will do to him after he touches the spin.
Assuming that $A$ measures the spin and finds it to be up, all that he can deduce is that the state of the spin
is now $\ket\uparrow$ and the total state is $\ket\uparrow\ket{\bar U}\ket{\rm B}$ (with $\ket{\bar U}$
the state where $A$ has observed the spin up and knows the measurement to which he will be subjected
afterwards, contrasted to state $\ket U$ without that knowledge, and similarly for $\ket{\bar D}$).
Accounting for the upcoming measurement on him, $A$ can deduce the evolution of state
{\small{\bea
&\hskip -0.1cm A~\text{{has~measured~spin~up~and~deduces~state~to~be}}~~~~~~~\Rightarrow ~~~&\ket\uparrow \ket{\bar U} \ket{\rm B} \label{Dog}\\
&\hskip -0.5cm A~\text{{deduces~state~to~become~after~his~cat~measurement}} \Rightarrow ~
&\hskip -0.5cm {\scriptstyle{1\over 2}} \bb
\ket\uparrow \bb\bigl[(\ket{\bar U}\bb+\bb \ket{\bar D}) \ket{\rm Y}+(\ket{\bar U} \bb-\bb \ket{\bar D}) \ket{\rm N}\bigr] ~~~~~~~
\eea}}
\noindent
$A$ can predict that he can observe the spin to be up after he has observed it to be down, but clearly missed the possibility
that he can observe the spin to be down after he has observed it to be up.
If the initial state was as in (\ref{Cat}), and $A$ measured the spin to be up and then concluded
that the spin will be measured to be up later on, as implied by (\ref{Dog}), he would have
50\% probability to be wrong. The lesson we draw is:
\vskip -0.1cm
\noindent
{\it Observation 2:} Observers cannot in general modify Born rules to fully account for cat measurements on themselves without
prior knowledge of the state of the full system and its later evolution.

I should stress that there is nothing unusual about cat states like $\ket U\bb \pm\bb \ket D$ {\it per se}:
it is only the possibility of directly measuring them that creates issues. By contrast, their indirect measurement
(deduction) poses
no problems. For example, consider the scenario where $\rm B$ knows the initial state of the full system
and the fact that $A$ will measure the spin in the $z$-basis, but now she does
not measure $A$; instead, she measures the spin in the $x$-basis $\ket\leftarrow$ and $\ket\rightarrow$.
The corresponding process would be
{\small{\bea
& \text{Initial~state} ~~~~~~~~~&~~~~\textstyle{1\over \sqrt 2} \bigl(\ket\uparrow + \ket\downarrow \bigr) \ket A \ket{\rm B} \label{Pet}
\\
~ & ~ \cr
&\hskip -0.5cm A~\text{measures~spin~in~{\it z}~axis}~\Rightarrow ~~~&~~\textstyle{1\over \sqrt 2} \bigl(\ket{\uparrow} \ket{U} + \ket{\downarrow} \ket{D} \bigr) \ket{\rm B}  \\
& ~~~         &=\textstyle{1\over 2} \Bigl[\ket\rightarrow\bb \bigl(\ket U\bb+\bb \ket D \bigr) +\ket\leftarrow\bb \bigl(\ket U \bb-\bb \ket D\bigr)\Bigr] \ket{\rm B} \cr
~ & ~ \cr
&\hskip -0.5cm {\rm B}~\text{measures~spin~in~{\it x}~axis}~\Rightarrow ~~&~\textstyle{1\over 2} 
\Bigl[\ket\rightarrow\bb \bigl(\ket U\bb+\bb \ket D \bigr)\ket{\rm R} +\ket\leftarrow\bb \bigl(\ket U \bb-\bb \ket D\bigr) \ket{\rm L} \Bigr]
\eea}}
$\rm B$ now knows that if she has seen the spin to point right ($\ket\rightarrow$) then $A$ is in the cat state 
$\ket U\bb+\bb\ket D$,
and similarly if she has seen it point left, so she has indirectly measured $A$ in a cat state (that is, she
has deduced by her knowledge of the state of the system and its evolution that $A$ is in a cat state).
However, this causes no problems: although now again $A$ has 50\%
probability to see the spin up or down, irrespective of what he observed before, he is not surprised, since the
spin was disturbed by ${\rm B}$'s measurement. Crucially, $A$ can use information on what $\rm B$ will
measure to make reliable predictions about later measurements based on his updated state after he observes
the spin, and without
knowledge of the full state before he makes a measurement; a repetition of the steps that led to equations
(\ref{reda}-\ref{redw}) would produce the same final outcomes.

Note that the above statements hold generically. In special situations with specific relations between cat and
non-cat measurements, and with observers having partial information on what measurements will
be performed, some predictability may be salvaged for them. To demonstrate this, consider the generalized
situation of eq. (\ref{Cat}) in which $A$ is measured in the new orthogonal cat states $\ket Y$, $\ket N$ and the
spin is in the state $\ket \chi$
\bea
&&\ket Y = a\ket U + b\ket D ~,~~~\ket N = b\ket U - a\ket D \label{YN} \\
&&\ket \chi = c \ket \uparrow + d \ket \downarrow ~~{\text {with}}~~ a^2 + b^2 = c^2 + d^2 = 1
\eea
(By choosing the phases of $\ket U$, $\ket D$, $\ket Y$ and $\ket N$ appropriately we can make
$a$ and $b$ real and positive, and similarly for $c$ and $d$ by choosing the phases of $\ket \uparrow$
and $\ket \downarrow$, since we will not measure the spin in any other basis in this setting.)
Following the same sequence of measurements as in (\ref{Cat}) we obtain the final state
{{\bea
& \text{Initial~state} ~~&~~~~ \ket{A}\bigl(c\ket{\uparrow} + d\ket{\downarrow} \bigr) \ket{\rm B} \\
~ &~ \cr
& \text{Final~state} ~~~
&~~\,\ket U\Bigl(a^2 c \ket\uparrow\bb \ket{\rm Y}\bb+\bb b^2 c\ket\uparrow \ket{\rm N} \bb+\bb abd
\ket\downarrow \bb\ket{\rm Y}\bb -\bb abd\ket\downarrow \ket{\rm N} \Bigr)\\
& ~~ &+\ket D \Bigl(abc
\ket\uparrow\bb \ket{\rm Y}\bb - \bb abc\ket\uparrow \ket{\rm N}\bb +\bb b^2 d
\ket\downarrow \bb\ket{\rm Y}\bb +\bb a^2 d\ket\downarrow \ket{\rm N} \Bigr)
\nonumber
\eea}}
\vskip -0.4cm
\noindent
The probabilities for the unexpected outcomes ``$A$ measures spin down given that he first measured it up" $p_{ud}$
and ``$A$ measures spin up given that he first measured it down" $p_{du}$ are
\be
p_{ud} = {2a^2 b^2 d^2 \over c^2 + 2a^2 b^2 (d^2 - c^2 )}~,~~~
p_{du} = {2a^2 b^2 c^2 \over d^2 + 2a^2 b^2 (c^2 - d^2 )}
\ee
These probabilities are maximized for $a=b=1/\sqrt 2$, the ``maximal" cat state, and
become $p_{ud} = d^2$, $p_{du} = c^2$, reproducing the result of eq. (\ref{Cat}) for $c=d=1/\sqrt 2$.
The state of maximal uncertainty for $A$ after having measured the spin up, $p_{ud} = p_{uu} = 1/2$,
arises for $c = ab\sqrt 2$, and similarly after having measured the spin down for $d = ab\sqrt 2$.

Based on any partial information that $A$ may possess on the initial state of the spin and his upcoming cat measurement,
$A$ may have some limited predictive power. E.g., if $A$ is informed that he will be measured in the exact same
superposition as the spin ($a=c, b=d$ or $a=d, b=c$), then he can deduce ``After I measure the spin, the probability
to find the opposite value in the subsequent measurement is less than $2/3$"; if $A$ is informed that he will be
measured in a state correlated with the spin state as in $c=ab\sqrt 2$, then he can deduce ``If I measure the spin and
find it up, the next measurement will be completely random; if I find it down, the next measurement is at least as
likely to find it down as it is to find it up" ($p_{ud} = 1/2, p_{du} \le 1/2$); etc.
However, no general rule emerges for estimating probabilities, and in the absence of any information on the
initial state of the spin and the upcoming cat measurement, $A$ is completely ignorant about the outcome of his
next spin measurement (both $p_{ud}$ and $p_{du}$ range from 0 to 1).

The above situation also highlights the distinction between cat measurements and ordinary measurements.
Observers do interact and ``measure'' each other continuously, but their interactions produce evolutions
of their conscious states and not superpositions of macroscopically distinct states. By contrast, the
cat measurement of eq. (\ref{YN}) is part of a continuum that interpolates between no cat measurement
($a=0$ or $b=0$) and the maximal cat measurement ($a=b=1/\sqrt 2$). As the cat measurement degenerates
($a\to 0$ or $b \to 0$) the probabilities of the surprising outcomes $p_{ud}$ and $p_{du}$ go to zero and the
standard Born rules are recovered: there is no ``discontinuous" loss of predictability. How such a continuous
evolution of states can be obtained with a cat measurement will be described later, when the execution and
feasibility of cat measurements are examined.

{{\bf Communication of information}}: The previous arguments apply to observer $A$'s prediction of experimental outcomes as experienced by himself.
It is also useful, and relevant for the FR thought experiment,
to examine how his predictions can be used by other observers; that is, how observers can communicate information.

Consider a third observer $C$ (call him Chris) who does not
participate in the measurements, nor is he going to be cat-measured himself, but derives conclusions based on information
from $A$. If observer $A$ directly communicates his prediction to $C$ about the value of the spin before he is cat-measured,
then clearly $C$
can treat this information as reliable. Such a communication amounts to entangling the cognitive states of $A$ and $C$,
and therefore of $C$ and the state of the system (spin) measured by $A$. It is, thus, indistinguishable from $C$ measuring
the spin himself. The subsequent cat measurement of $A$ affects neither $C$ nor the spin, and in the absence of cat measurements on himself, $C$ can make reliable predictions.

The situation is similar with indirect (deduced) measurements, that is, for states where cognitive states of $A$ and $C$
become entangled as a result of the dynamical evolution of the system without direct communication between them.
If $C$ can reliably deduce such an entanglement from his knowledge of the system, he can treat the information deduced
from $A$'s measurement (unreliable for $A$ himself) as reliable. A simple example is the evolution of a state involving
two entangled spins and observers $A$, {\rm B}, and $C$:
{\small{\bea
& \text{Initial~state} ~~~~~~~~~&~~~~\textstyle{1\over \sqrt 2} \bigl(\ket{\uparrow \, \rightarrow} + \ket{\downarrow\, \leftarrow} \bigr) \ket{A}  \ket{C} \ket{\rm B} \label{Indirect}\\
~ &~ \cr
&\hskip -0.8cm A~\text{measures~first~spin~along~}z~\Rightarrow ~~~&~~\textstyle{1\over \sqrt 2} \Bigl(\ket{\uparrow\,\rightarrow}
\ket{U} + \ket{\downarrow\,\leftarrow} \ket{D} \Bigr) \ket{C} \ket{\rm B} \\
~ &~ \cr
&\hskip -0.8cm C~\text{measures~second~spin~along~}x~\Rightarrow ~~~&~~\textstyle{1\over \sqrt 2} 
\Bigl(\ket{\uparrow\,\rightarrow} \ket{U} \ket{R} + \ket{\downarrow\,\leftarrow} \ket{D} \ket{L} \Bigr) \ket{\rm B} \\
~ &~ \cr
&\hskip -1.1cm {\rm B}~\text{measures~{\it A}~in~cat~state}~\bb\Rightarrow &\hskip -0.0cm \textstyle{1\over \sqrt 8} 
\Bigl\{ \ket{\uparrow\,\rightarrow} \bigl[(\ket U\bb\bb+\bb\bb \ket D)\bb \ket{\rm Y}\bb+\bb(\ket U \bb\bb-\bb\bb \ket D)\bb \ket{\rm N}\bigr]\ket{R} \\
&~~~ & ~~+ \ket{\downarrow\,\leftarrow} \bigl[ (\ket U\bb\bb+\bb\bb \ket D)\bb\ket {\rm Y}\bb -\bb (\ket U\bb\bb-\bb\bb \ket D) \bb\ket{\rm N} \bigr] \ket{L}\Bigr\}\nn
\eea}}
Although $A$ and $C$ never directly interact, the knowledge by $C$ that their states are entangled after $C$'s
measurement of the second spin is enough for $C$ to correctly predict the result of a measurement of the first
spin, even after $A$ is cat-measured. (Note that the last two measurements commute: performing them in the opposite
order changes neither the final state nor the deductions of $C$ and $\rm B$.)

Things become trickier, however, when $C$ is himself going to be cat-measured. The previous conclusions
about predicting or communicating results on a later measurement still hold.
However, if the measurement in question is a cat measurement that involves himself,
$C$ can neither make reliable predictions, nor transmit reliable information, either directly or indirectly.

Direct transmission of information by $C$ is immediately excluded: this would
entangle his state with that of another observer, which would disturb the measured system (himself). What is
subtler is the fact that even indirect transmission of information, which would not disturb him, is unreliable.
To demonstrate this, consider the process involving observers $A$ (in states $\ket U$ and $\ket D$) and $C$
(in states $\ket L$ and $\ket R$) in an entangled state, with {\rm B} measuring them in cat states
$\ket U\bb +\bb \ket D$ and $\ket L \bb+\bb \ket R$, and with all observers knowing the full initial state
of the system. The state evolution is:
{\small{\bea
& \text{Initial~state} ~~~~~~~~&~~~\textstyle{1\over \sqrt 3} \bigl(\ket{U} \ket{L} + \ket{D} \ket{L} +\ket D \ket{R} \bigr) \ket{\rm B}\\
~ &~ \cr
&\hskip -0.9cm ~\text{{\rm B}~cat-measures~}A~\bb\bb\Rightarrow &\hskip -0.2cm \textstyle{1\over 2\sqrt 3} \Bigl[ 
2\bigl(\ket U \bb+\bb \ket D \bigr) \ket{\rm Y} \ket{L}\bb +\bb \bigl(\ket U + \ket D \bigr) \ket{\rm Y}\ket{R}
\bb-\bb \bigl(\ket{U} \bb-\bb \ket{D} \bigr) \ket{\rm N} \ket R \Bigr] \\
~ &~ \cr
&\hskip -0.9cm {\rm B}~\text{cat-measures~}C~\bb\bb\Rightarrow &\hskip -0.1cm \textstyle{1\over 4\sqrt 3} 
\Bigl[ 3\bigl(\ket U \bb\bb+\bb\bb \ket D \bigr) \bigl(\ket L \bb\bb+\bb\bb \ket R \bigr) \ket{\rm YY} 
+ \bigl(\ket U \bb\bb+\bb\bb \ket D \bigr) \bigl(\ket L \bb\bb-\bb\bb \ket R \bigr) \ket{\rm YN}  \label{NoWay}\\
&~~~ & ~~~~ - \bigl(\ket U \bb\bb-\bb\bb \ket D \bigr) \bigl(\ket L \bb\bb+\bb\bb \ket R \bigr) \ket{\rm NY} 
+ \bigl(\ket U \bb\bb-\bb\bb \ket D \bigr) \bigl(\ket L \bb\bb-\bb\bb \ket R \bigr) \ket{\rm NN} \Bigr] \nn
\eea}}
In the initial state, $A$ in state $\ket D$ is entangled with state $\ket L \bb+\bb \ket R$ of $C$.
Since he knows the initial state of the system, he can deduce this entanglement and he predicts
the result Yes for the cat-measurement on $C$. This prediction is invalid for $A$ himself,
in view of his later cat measurement, but can
be reliably passed to other observers. $C$ in the state $\ket R$ is entangled with $\ket D$,
so $C$ in that state can inherit the conclusion of $A$ in $\ket D$ and indirectly
conclude that the result of his own cat measurement will be Yes. This prediction is invalid for $C$ himself,
in view of his own cat measurement, but could presumably be reliably passed to other observers.

After {\rm B} cat-measures $A$, her state $\ket{\rm N}$ is entangled with state $\ket R$ of $C$, so if indirect transmission
of information from $C$ were reliable, {\rm B} in state $\ket{\rm N}$ would conclude that a measurement of $C$ would yield Yes.
Yet this prediction is invalidated by the last state in (\ref{NoWay}), which includes the state $\ket {\rm NN}$ in which {\rm B},
originally in the state $\ket{\rm N}$, obtains the result No for the cat measurement of $C$.

The lesson we draw from the above chain of arguments is:
\vskip -0.1cm
\noindent
{\it Observation 3:} Observers cannot, in general, relate reliable information to other observers if
both observers are going to be subject to cat measurements.

The previous arguments are also relevant to the operational validity of the assumption of ``state collapse."
Taking, e.g., the setup described in (\ref{Dog}), what observer $A$ is doing is essentially state collapse: based on the
information that he obtained from his measurement of the spin, he assumes the state to be an eigenstate of this measurement.
This is the best that he can do, lacking any independent knowledge of the full state before making any observations, and that's what we usually do after measurements,
and in general we get away with it: cognitive states corresponding to other possible outcomes do not
interfere, and results drawn upon the reduced state by an observer in that state are valid.
As demonstrated in eqs.~(\ref{reda}-\ref{redw}), cat measurements change that:
by mixing macroscopic states of the observer they make alternatives interfere, and wavefunction collapse
yields unreliable results. A restatement of the first lesson of the paper would be:
\vskip -0.1cm
\noindent
{\it Observation 4:} Observers cannot use state collapse if they will be cat-measured.

The above considerations demonstrate that the deduction rules of quantum mechanics do not hold if the observer suffers
cat measurements, and need to be supplemented with the condition of absence of such measurements. This
lifts the paradox obtained by FR without modifying the essence of quantum mechanics, as I will demonstrate.

{{\bf FR's paradox}}: The FR thought experiment involves two agents F and 
$\bar{\rm F}$ and two Wigner friends W and $\bar {\rm W}$ in a chain of events and measurements.
F and $\bar{\rm F}$ can measure two spins (one of them viewed as a ``dice"), while W and $\bar{\rm W}$
can perform cat measurements on F and $\bar{\rm F}$ themselves, labeling the results of these
measurements ``ok" or ``fail." This augmented setup is needed to
produce a set of conclusions, derived by consistency between the quantum mechanical predictions of
the various observers, that lead to a contradiction. The chain of events and conclusions of the various agents and their mutual interrelation in their temporal succession, as in Table 3 in FR's paper, are summarized below ($\ket{h}$ and $\ket{t}$ are the states of the ``dice" spin):

0. The initial state of the two spins is $\bigl(\ket{h} \ket{\downarrow}\bb+\bb\ket{t}\ket{\downarrow}\bb+\bb\ket{t}\ket{\uparrow}\bigr)/$
{\small \bb\bb\bb$\sqrt3$}.

1. Agent $\bar{\rm F}$ measures the dice spin, finds it to be $\ket{t}$ and becomes certain that agent W will obtain the outcome ``fail" at the final measurement of the experiment (statement $\bar{\rm F}^{n:02}$ in FR)

2. Then agent F performs a measurement of the second spin, finds it to be $\ket{\uparrow}$, becomes certain that $\bar{\rm F}$
observed the dice to be $\ket{t}$, and becomes certain that
W will obtain the outcome ``fail" at the end because agent $\bar{\rm F}$ is certain of this outcome (statement
${\rm F}^{n:14}$)

3. Then agent $\bar{\rm W}$ performs a cat measurement on $\bar {\rm F}$'s lab, finds her to be in the $\overline{\rm ok}$ state,
becomes certain that F observed the spin $\ket{\uparrow}$ and becomes certain that W will obtain the outcome ``fail" because
agent F is certain of this outcome (statement ${\bar{\rm W}}^{n:24}$)

4. Finally, agent W becomes certain he will obtain the outcome ``fail" because agent $\bar{\rm W}$ is certain of
this outcome (statement ${\rm W}^{n:28}$),  but subsequently measures F and obtains the result ``ok", leading to a contradiction

Statement $\bar{\rm F}^{n:02}$ by agent $\bar{\rm F}$ is a prediction based on the application of standard Born inference
rules on the specific state $\ket{t}$ that $\bar{\rm F}$ obtains after measuring the dice. Each of the remaining statements 2, 3
and 4 relies on the validity of drawing conclusions based on the previous statement.

FR include the standard quantum mechanical inference rule as one of their basic assumptions (Assumption $Q$). In fact,
FR used a weaker, non-probabilistic quantum mechanical rule, applicable to eigenstates of the observed quantity,
which was sufficient for the prediction of agent $\bar{\rm F}$ and the derivation of their result. I state their assumption
below, slightly paraphrased and in Schr\"odinger language:

{\it Assumption Q:} If an agent {\rm A} has established at time $t_0$ that a quantum system S is in a state that
will evolve at time $t$ into an eigenstate of an observable X with eigenvalue $\xi$, then agent {\rm A} can conclude:
``I am certain that $X = \xi$ at time t.''

With the additional condition implied by the considerations in the present work, this assumption should be modified as:

{\it Assumption $Q'$:} If an agent {\rm A} has established at time $t_0$ that a quantum system S is in a state that
will evolve at time $t$ into an eigenstate of an observable X with eigenvalue $\xi$, and if A knows that
no cat measurements will be performed on A during the interval $(t_0,t)$, then agent {\rm A} can conclude:
``I am certain that $X = \xi$ at time t.''

The other assumption of FR is consistency between the predictions of different observers (Assumption $C$).
I state their assumption below, again paraphrased in Schr\"odinger language:

{\it Assumption C:} If an agent {\rm A} has established at time $t_0$  that another agent {\rm B}, reasoning according
to quantum mechanics, is certain that an observable X will have the value $\xi$ at time $t$, then agent {\rm A} can
conclude: ``I am certain that $X=\xi$ at time t."

With the additional condition implied by the considerations in the present work, it should be modified as:

{\it Assumption $C'$:} If an agent {\rm A} has established at time $t_0$  that another agent {\rm B}, reasoning according
to quantum mechanics, is certain that an observable X will have the value $\xi$ at time $t$, and if A knows that
no cat measurements will be performed on either A or B during the interval $[t_0,\hb t]$, then agent {\rm A} can conclude: \bb
``I am certain that $X\bb\hb=\hb\xi$ at time t."

The third assumption of FR (Assumption S) is that of logical consistency, precluding the derivation of
mutually incompatible results, and is not (and should not be!) modified.

With the assumptions thus modified, FR's argument can stumble at a couple of steps: agent
$\bar {\rm F}$ draws her conclusion about the measurement output of agent W at the final step of the experiment
based on her present state, leading to statement 1.
However, this conclusion is invalidated by the fact that $\bar {\rm F}$ will be herself cat-measured 
before that final step, as per Assumption $Q$.

Still, this is not necessarily fatal for FR's argument,
since $\bar {\rm F}$'s conclusion could possibly be communicated reliably to another agent,
leading to statement 2. However, this conclusion is invalidated by the fact that F, who receives this
conclusion, will also be cat-measured, triggering the caveat of
assumption C. From that point, agent F cannot communicate reliable information to any other agent.
Statements 3 and 4 cannot be derived, and no contradiction ensues.

Note that the above argument is valid even if agent
$\bar {\rm F}$ is ``destroyed" after making the prediction and being measured
by $\bar {\rm W}$, as scripted in some scenaria, since $\bar {\rm F}$ does not participate in any of the remaining
measurements or deductions. Unitarity forbids the destruction of $\bar {\rm F}$ into a universal destroyed
state: each orthogonal state of $\bar {\rm F}$ will be destroyed into distinct orthogonal states,
which serve as proxies for the undestroyed states of $\bar {\rm F}$ until the end of the thought experiment,
replicating essentially the same state evolution.

This analysis highlights the ingenuity of the thought experiment proposed by FR:
a contradiction in quantum theory could easily
have been obtained by a simple scenario such as the one of equ.~(\ref{Cat}), with agent $A$ making a prediction for the
measurement of the spin after he has measured it once, and seing it invalidated in his subsequent measurement after
his own cat measurement. However, FR wanted the contradiction to be obtained by an agent not suffering
himself a cat measurement, thus requiring an indirect transfer of information. Yet such information transfers, as I argued
around equ.~(\ref{Indirect}), are often reliable. A situation with an unreliable transfer of information was needed,
necessitating a second cat-measured agent as well as non-cat-measured agents. In fact, the situation
in FR's thought experiment exactly parallels the one in equ.~(\ref{NoWay}), with $\bar {\rm F}$
and F playing the role of
$A$ and $C$, and {\rm B} subsuming the roles of
$\bar {\rm W}$ and W, while the dice and spin serve to produce the appropriate initial entangled state.
Overall, FR's setup is useful in sharpening our intuition and alerting us to the limitations of predictions and communication
between observers that suffer cat measurements.

Other arguments for lifting FR's paradox have been offered, and the difficulties caused by cat measurements
have been highlighted, with statements such as Scott Aronson's witty aphorism ``It's hard to think when
someone Hadamards your brain" \cite{Aron}, or Lenny Susskind's comment in Renato Renner's seminar
\cite{talk} about 
``closed loops" in the many-world interpretation. My arguments sharpen the issue into precise statements
and propose specific modifications of Assumptions $Q$ and $C$. They also eliminate the possibility of a general modification
of quantum rules (based only on observationally available data) to take into account cat measurements,
or at least show that the quantitative rules for such modifications are nontrivial and as yet to be formulated.
As I stated early in the paper, I prefer to eschew the many-worlds view as it offers no
conceptual advantages
in the presence of cat measurements, since the question of ``who can branch the
world?" is essentially equivalent to ``who can collapse wavefunctions?".

Finally, it should be obvious why cat measurements are necessary to produce FR's paradox while classical measurements
would not do it. An agent could make a
prediction and reliably relate it to another agent before getting confused by a classical
“bang on the head,” producing no inconsistencies. By contrast, the information deduced from two cat-measured observers
can become unreliable as their
states are scrambled after their cat measurements, which is a pure quantum effect. This is
the essence of eq.~(\ref{NoWay}) and of FR's thought experiment, and this is what the modified
Assumption C warns about.

{{\bf Are cat measurements possible?}} The possibility (or suspicion) of cat measurements performed upon ourselves or our experimental apparati would
be catastrophic for our ability to usefully apply quantum theory. The success of quantum mechanics in every context where
it was applied so far is evidence that such measurements are either physically impossible or of vanishingly small probability.
Observers, of course, do interact and ``measure'' each other continuously, but their interactions are
essentially classical, that is, they never create superpositions of macroscopically distinct states.

The von Neuman realization of a cat measurement on $A$ would require
coupling the measured system with the momentum of the position operator of the
``needle'' of the observation apparatus. Such an interaction for the process (\ref{Cat}) would be
\be
h_{_I} = \lambda\, p\, \Pi
\ee
with $\lambda$ a real coupling constant, $p$ the momentum operator dual to the position $x$ of the needle of a measuring 
apparatus in observer $\rm B$'s lab, and $\Pi$ an operator with $\ket U\bbb+\bbb \ket D$ and $\ket U\bbb-\bbb \ket D$ as
non-degenerate eigenstates. Up to irrelevant additive and multiplicative constants, such an operator can be
expressed as
\be
\Pi = \ket U \bra D + \ket D \bra U
\ee
and would act as
\be
\Pi\, \ket{U} = \ket{D} ~,~~~ \Pi\, \ket{D} = \ket{U}
\ee
That is, $\Pi$ is an exchange operator that acts on $A$ and changes his state from one where he has
observed the spin to be up to one where he has observed it to be down, and {vice versa}. 
Applying the Hamiltonian $h_{_I}$ for time $t$ produces the unitary evolution
\be
{\rm U} = e^{-i h_{_{\bb \small I}} t} = \half e^{-i\lambda tp} (1+\Pi ) + \half e^{i\lambda tp} (1-\Pi)
\ee
Starting with the initial state $\ket  U \ket{\rm B}$, the state at time $t$ woul be
\be
{\rm U} \ket U \ket{{\rm B}} =
\half (\ket U + \ket D ) \ket{{\rm B} (\lambda t)} + \half (\ket U - \ket D ) \ket{{\rm B} (-\lambda t)}
\label{meast}\ee
where $\ket{{\rm B} (x)}$ represents the state of $\rm B$ with her measuring device needle's position
shifted by $x$ (the initial state of $\rm B$ would be $\ket{\rm B} = \ket{{\rm B} (0)}$). If the initial
uncertainty in the position of the needle is $\delta$, then after time $T > \delta / \lambda$, $\rm B$
could decide with certainty if the needle moved in the positive or negative direction, and at that
time $\ket{{\rm B} (\lambda T)} = \ket{\rm Y}$ and $\ket{{\rm B} (-\lambda T)} = \ket{\rm N}$,
leading to the final state
\be
\half (\ket U + \ket D ) \ket{{\rm Y}} + \half (\ket U - \ket D ) \ket{{\rm N}}
\ee
The important fact is that the state in (\ref{meast}) never contains a state of confusion for
$A$; it is always a superposition of $\ket U$ and $\ket D$. Both states $\ket U$ and $\ket D$ 
are undisturbed states of clear certainty about the value
of the spin (up or down) and the full state at all times is an eigenstates of $A$'s ``certainty operator."
This justifies the statement that $A$ would feel absolutely nothing during a ``clean" measurement such as
the one above and would not even be aware that he is cat-measured. It also demonstrates the asymmetry
in the situation: only $A$ suffers the action of exchange
operators. Ironically, $A$ never leaves a state of certainty, while $\rm B$ goes through a continuous set
of states of uncertainty $\ket{{\rm B} (\lambda t)}$ until she reaches her final state of certainty about
the outcome of the measurement.

Are cat measurements such as the one above physically realizable? In fact, 
the physics of performing cat measurements is prohibitive. 
Exchange operators are strongly nonlocal (essentially effecting ``teleportation") and hard to realize,
even in the simplest of systems. 
For example, the parity operator $P$ reflecting the position and momentum of a particle on the line,
can be realized as
\be
P = \exp\left[{i{\pi\over 2} \left(a x^2 + {p^2 \over a\hbar^2} - 1\right)}\right]~,~~
P x = -x P ~,~~ Pp = - p P
\ee
with $a$ a nonzero real constant (despite appearances, $P$ is both Hermitian and unitary).
This is a highly unphysical operator, involving an infinite sequence of local operators
(upon Taylor-expanding the exponential). Physical interactions are local, and no finite sequence of them
would reproduce $P$.

The realization of $\Pi$ would similarly involve nonlocal operators acting on the
macroscopically large number of particles making up observer $A$, and in a
highly coordinated pattern, pushing it outside the realm of physical possibilities. Interactions between observers,
however intense or even violent, are a collection of local individual interactions and will never reproduce $\Pi$. Even
a reasonable approximation of $\Pi$ would need to involve an exceedingly long sequence of operations whose execution would
require a time likely exceeding the lifetime of the universe.

Nevertheless, the question of whether cat measurements are in principle realizable is an interesting one and
remains essentially open.
I offered some arguments why such measurements would be practically impossible, but at the conceptual level it would be desirable
to have a proof of their full impossibility, perhaps involving locality, relativity, quantum field theory
(which does not even contain strictly factorizable, unentangled states of finite energy) or other physical principles. In fact, making the question a
meaningful one would require thermodynamics to enter the argument at some level.
Just as there is no sharp distinction between ``small" (quantum) and ``large"
(classical) systems, what constitutes an observer, and thus what is a cat measurement, is equally fuzzy.
It is often argued (or conjectured) that the arrow of time and the manifestation of consciousness are related to
entropy flow. In that case, a physically meaningful definition of cat measurements would necessarily involve large systems out of
equilibrium. The physical realization of operators like $\Pi$ could then possibly be excluded by entropic considerations.

In conclusion, quantum mechanics is alive and well, still challenging us to understand it to our intellectual and emotional
satisfaction. If cat measurements can be ruled out, quantum mechanics will become more
reliably predictive. If not, cat measurements will remain in our intellectual playground and may lead to interesting
and weird effects, and possibly new insights, although not to inconsistencies. I am biased for the former, but otherwise remain agnostic.

\vskip 0.4cm
{\bf Data Availability Statement}: Data sharing not applicable to this article as no datasets were generated or analysed during the current study.

\vskip 0.4cm
\noindent
{\bf Acknowledgements}:
I would like to thank Stuart Samuel for interesting me in the work of Frauchiger \& Renner and for relating his own,
different, explanation of the paradox, and Parameswaran Nair for a critical assessment of my argument. I
am especially thankful to Renato Renner for a useful correspondence and for sharing his insights.

\noindent
This work was supported by NSF under grant NSF-PHY-2112729 and by PSC-CUNY grants 65109-00 53
and 6D136-00 02.

\vskip 0.4cm
\noindent
{\bf Author contribution}: A.P. is the sole contributor to this work and is fully responsible for its contents.

\vskip 0.4cm
\noindent
{\bf Competing interests}: The author declares no competing interests.

\end{document}